\documentclass[runningheads]{llncs}
\usepackage[T1]{fontenc}
\usepackage{authblk}

\usepackage[dvipsnames]{xcolor}
\usepackage{graphicx}
\usepackage{tabularx}
\usepackage{paralist}
\usepackage{dirtytalk}
\usepackage{amsmath}
\usepackage{bbm}
\usepackage{amsfonts}
\usepackage{leftindex}
\usepackage{appendix}
\usepackage{changepage}

\DeclareMathOperator{\EX}{\mathbb{E}}

\begin{document}

\title{Approximation Algorithms for the  $b$-Matching and List-Restricted Variants of MaxQAP}
\titlerunning{$b$-Matching and List-Restricted Variants of MaxQAP}

\author{Jiratchaphat Nanta\inst{1} \and
Vorapong Suppakitpaisarn\inst{2} \and
Piyashat Sripratak\inst{1}}
\authorrunning{J. Nanta et al.}
%
\institute{Chiang Mai University, Thailand \and
The University of Tokyo, Japan}

\maketitle

\begin{abstract}
We study approximation algorithms for two natural generalizations of the Maximum Quadratic Assignment Problem (MaxQAP). In the \emph{Maximum List-Restricted Quadratic Assignment Problem}, each node in one partite set may only be matched to nodes from a prescribed list. For instances on $n$ nodes where every list has size at least $n - k$, we design a randomized $O(\sqrt{n}+k)$-approximation algorithm based on the linear-programming relaxation and randomized rounding framework of Makarychev, Manokaran, and Sviridenko. In the \emph{Maximum Quadratic $b$-Matching Assignment Problem}, we seek a $b$-matching that maximizes the MaxQAP objective. We refine the standard MaxQAP relaxation and combine randomized rounding over $b$ independent iterations with a polynomial-time algorithm for maximum-weight $b$-matching problem to obtain an $O(\sqrt{bn})$-approximation. When $b$ is constant and all lists have size $n - O(\sqrt{n})$, our guarantees asymptotically match the best known approximation factor for MaxQAP, yielding the first approximation algorithms for these two variants.
\end{abstract}

\section{Introduction}

Given two complete weighted graphs
$G = (V_G, E_G, w_G: E_G \rightarrow \mathbb{R}_{\geq 0})$ and 
$H = (V_H, E_H, w_H : E_H \rightarrow \mathbb{R}_{\geq 0})$
such that $|V_G| = |V_H| = n$, the \emph{Quadratic Assignment Problem} (QAP) is to find a bijection
$\pi: V_G \rightarrow V_H$ that optimizes the quadratic objective
$\sum_{u,v \in V_G} w_G(u,v)\, w_H(\pi(u), \pi(v))$.

In the classical minimization form, the goal is to minimize this sum. The QAP is a fundamental
combinatorial optimization problem that models the assignment of a set of facilities to a set of
locations with the aim of minimizing the total cost, which depends on the flow between facilities and
the distance between locations~\cite{koenig1957}. It is known to be NP-hard and is considered one of
the most challenging problems in the field of optimization~\cite{pardalos1994}. Numerous heuristic
and approximation algorithms have been proposed to address its computational complexity in practical
applications such as facility layout, scheduling, and electronics design~\cite{cela1998}. The QAP
has also recently been considered as a target for the Quantum Approximate Optimization Algorithm
(QAOA), a quantum heuristic designed for tackling hard combinatorial optimization
problems~\cite{codognet2022quantum,ye2023towards}. Its relevance to quantum algorithm design has
drawn increasing attention from the quantum computing community.

\noindent\textbf{MaxQAP and applications.}
While the standard QAP is typically formulated as a minimization problem, in this paper we focus on
its maximization variant, known as \emph{MaxQAP}. 
MaxQAP is also referred to as the graph matching problem in the fields of pattern recognition and
computer vision, where it has numerous applications~\cite{bougleux2017graph,leordeanu2005spectral,zhou2015factorized}.
Moreover, MaxQAP generalizes the graph isomorphism problem~\cite{grohe2018graph}. Owing to its broad
range of applications, several approximation algorithms have been proposed for MaxQAP~\cite{makarychev2014maximum,nagarajan2009maximum}.
The best known approximation ratio for the problem to date is $O(\sqrt{n})$.

\noindent\textbf{Motivation for constrained variants.}
In many applications, the assumption that we are free to choose an arbitrary bijection
$\pi:V_G \to V_H$ is overly idealized. 
    In pattern recognition, a landmark in one image may only correspond to landmarks in a specific
    region of another image. In network alignment, vertex attributes or metadata may restrict which
    node in one social network can plausibly correspond to a node in another. In quantum circuit
    placement, hardware calibration or architectural restrictions
    may forbid certain logical-to-physical qubit mappings. 
    
    Also, in noisy graphs or heterogeneous systems, enforcing a strict one-to-one correspondence can be
    too rigid. Nodes may represent facilities or resources with limited capacities, supernodes
    that aggregate several entities, or duplicated objects. Allowing a node to be matched multiple
    times can yield similarity measures that are more robust to outliers and local noise.

These considerations motivate two natural generalizations of MaxQAP that we study in this paper:
a \emph{list-restricted} variant and a \emph{$b$-matching} variant.

\noindent\textbf{List-restricted MaxQAP.}
The list-restricted graph isomorphism problem has been studied in settings where each node in graph
$G$ can be matched only to a node in a specified subset of $V_H$, and vice
versa~\cite{lubiw1981some}. This constraint reflects practical scenarios in which prior knowledge
rules out certain node correspondences, and has been investigated from a theoretical perspective in
the context of graph isomorphism~\cite{klavik2021graph}. In the \emph{list-restricted MaxQAP}, we
are given a list $\mathcal{L}(u)$ of admissible images for each $u$, and we seek a
matching $\pi$ such that $u$ is matched with a node in the list. Despite the similarity to list-restricted
graph isomorphism, to the best of our knowledge, no theoretical results currently exist for MaxQAP
under list restriction constraints, even in restricted regimes.

\noindent\textbf{MaxQAP with $b$-matching.}
A generalization of the classical maximum matching problem is the maximum $b$-matching problem,
where each node is allowed to be matched with up to $b$ other nodes instead of just one. This
extension captures a broader range of applications~\cite{vazirani2022online} while maintaining
tractability, as the problem remains solvable in polynomial time~\cite{10.1145/800061.808776,kleinschmidt1995strongly}.
In our setting, we replace the bijection $\pi$ by a $b$-matching between $V_G$ and $V_H$, and
evaluate the same quadratic objective. While solutions to MaxQAP reflect the overall similarity
between the input graphs $G$ and $H$, the $b$-matching variant offers greater robustness by
providing more accurate similarity measures in the presence of corrupted nodes or inaccurate edge
weights.

The $b$-matching variant of the Maximum Edge-Pair Embedding Bipartite Matching problem~\cite{nguyen2021maximum},
which shares a strong connection with MaxQAP, has been explored in~\cite{buahong2025finding}. In
contrast, to the best of our knowledge, no existing work has extended the MaxQAP formulation itself
to incorporate a $b$-matching constraint.

\noindent\textbf{Relation to GQAP.}
The Generalized Quadratic Assignment Problem (GQAP) has been the subject of various
studies~\cite{lee2004generalized,mckendall2017heuristics}. These works typically focus on selecting
a subgraph of the input bipartite graph in which nodes on one side have a degree constraint of
exactly one, while nodes on the other side may be assigned to multiple partners. This formulation
differs from the $b$-matching setting, where the degrees of \emph{all} nodes are bounded by $b$. To
the best of our knowledge, approximation algorithms for GQAP have not yet been studied in the
literature, and existing work is largely heuristic.

\subsection{Our Contributions}

In this work we initiate the approximation-theoretic study of list-restricted and $b$-matching
variants of MaxQAP. Our main contributions are threefold:
\begin{enumerate}
    \item We formalize the list-restricted and $b$-matching variants of the MaxQAP in
    Section~\ref{sec:problem}, providing a unified framework that captures both types of
    constraints.

    \item Building on the randomized rounding technique introduced in~\cite{makarychev2014maximum},
    in Section~\ref{sec:list-restricted}, we develop an $O(\sqrt{n}+k)$-approximation algorithm for
    the list-restricted variant of MaxQAP when each node $u \in V_G$ can be
    matched to a node in a subset $\mathcal{L}(u)$ of size at least $|\mathcal{L}(u)| = n - k$.

    \item In Section~\ref{sec:QbAP}, we extend the randomized rounding technique to the
    $b$-matching setting and obtain an $O(\sqrt{bn})$-approximation algorithm for the $b$-matching
    variant of MaxQAP. 
\end{enumerate}
In particular, for the List-Restricted MaxQAP when $k = O(\sqrt{n})$ and for the MaxQbAP when $b$ is a constant, our approximation guarantees match
    the best known $O(\sqrt{n})$ approximation ratio for standard MaxQAP up to constant factors.

\subsection{Technical Overview}

The $O(\sqrt{n})$-approximation for the MaxQAP in \cite{makarychev2014maximum} is obtained via randomized rounding of a linear-programming (LP) relaxation. They prove that the rounding achieves, in expectation, an $\mathrm{\Omega}\bigl(1/\sqrt{n}\bigr)$ fraction of the LP optimum. For the analysis, the LP objective is partitioned into a \emph{heavy} and a \emph{light} part: for each node $p\in V_H$, let $\mathcal{W}_p$ be the set of the $\lceil \sqrt{n} \rceil$ nodes $q\in V_H$ whose weights $w_H(p,q)$ are the largest; the heavy part consists of contributions with $q\in\mathcal{W}_p$, and the light part is the remainder. The rounding guarantees that the result objective is  $\mathrm{\Omega}\bigl(|\mathcal{W}_p|/n\bigr) = \mathrm{\Omega}\bigl(1/\sqrt{n}\bigr)$ fraction of the light part in the optimal solution, yielding the stated $O(\sqrt{n})$ approximation ratio. 

To obtain an algorithm for the list-restricted MaxQAP, we modify the LP relaxation to satisfy the list-restricted constraint. This
modification can substantially reduce our objective value, because some nodes in
\(\mathcal{W}_p\) cannot contribute to the randomized rounding objective under the
new constraints. As a result, the stated fraction of the light part is no longer guaranteed to be achieved. To solve this issue, suppose each node in \(G\) can be matched to at least
\(n - k\) nodes for a constant \(k\). We expand \(\mathcal{W}_p\) to have
size \(\lceil (\sqrt{n+k^2}+k)/2 \rceil\). Even if up to \(k\) of the top choices are excluded by
the lists, at least \(\lceil(\sqrt{n+k^2}-k)/2\rceil\) candidates remain. Hence, the rounding still captures
an \(\mathrm{\Omega}((\sqrt{n+k^2}-k)/2n) = \mathrm{\Omega}(1/\sqrt{n}+k)\) fraction of the LP optimum, yielding an
\(O(\sqrt{n}+k)\)-approximation for the list-restricted problem.



Our approach for the $b$-matching variant of MaxQAP leverages the fact that a \(b\)-matching can be decomposed into \(b\) disjoint matchings. We therefore perform \(b\) iterative rounds of the randomized rounding process, then
combine the results obtained from those $b$ rounds. While a straightforward analysis of this approach yields an $O(b\sqrt{n})$-approximation, our refined analysis improves the guarantee to an $O(\sqrt{bn})$-approximation.


\section{Problem Formulation}
\label{sec:problem}
\textbf{List-Restricted MaxQAP.}
An instance of the maximum list-restricted quadra-tic assignment problem (List-Restricted MaxQAP) consists of two weighted graphs, $G = (V_G,E_G,w_G)$ and $H = (V_H,E_H,w_H)$ such that $|V_G| = |V_H| = n$, and a nonempty restricted list $\mathcal{L}(u) \subseteq V_H$ for each node $u \in V_G$. We also denote $k = \max_{u \in V_G} (n - |\mathcal{L}(u)|)$. The set of feasible solutions consists of \textit{compatible matchings} which are matchings on a bipartite graph $(V_G,V_H,E_{GH})$ where $E_{GH} = \{ (u,p) \in V_G \times V_H : p \in \mathcal{L}(u)\}$. The objective is to find a compatible matching $M$ that maximizes $\sum\limits_{(u,p),(v,q) \in M} w_G(u,v)w_H(p,q).$

\noindent \textbf{MaxQbAP.}
Suppose that we have two weighted graphs, $G = (V_G,E_G,w_G)$ and $H = (V_H,E_H,w_H)$ such that $|V_G| = |V_H| = n$, and a positive integer $b \leq n$. The Maximum Quadratic $b$-matching Assignment Problem (MaxQbAP) is to find a $b$-matching $^b\!M$ between $V_G$ and $V_H$ that maximizes 
\[
\sum_{u,v\in V_G}\sum_{p,q\in V_H}
  w_G(u,v)\,w_H(p,q)\;
  \mathbf{1}\!\left(
    \{(u,p),(v,q)\}\subseteq {}^{b}\!M \;\lor\;
    \{(u,q),(v,p)\}\subseteq {}^{b}\!M
  \right)
\]

The objective uses an indicator to check whether an edge $\{u,v\}\in E_G$ is mapped to $\{p,q\}\in E_H$ by ${}^{b}\!M$. Because this form is cumbersome to analyze, we instead study a variant, \textbf{dup-MaxQbAP}, which seeks a $b$-matching ${}^{b}\!M$ that maximizes
$\sum\limits_{(u,p),(v,q)\in{}^{b}\!M} w_G(u,v)\,w_H(p,q)$. If $(u,p),(u,q),(v,p),(v,q) \in {}^{b}\!M$, then the term $w_G(u,v)\,w_H(p,q)$ is counted twice in the objective of dup-MaxQbAP, whereas in our extension it should be counted only once. 

In Appendix~\ref{appendix:dup-maxqbap}, we show that any
\(\alpha\)-approximation for dup-MaxQbAP induces a
\(2\alpha\)-approximation for MaxQbAP. Hence an
\(O(\sqrt{bn})\)-approximation for dup-MaxQbAP also holds for
MaxQbAP, and we therefore focus on the dup-MaxQbAP formulation for
the remainder of the paper.

For convenience, we use a symbol \(\perp\), distinct from every
\(u \in V_G\), to represent an unmatched outcome. Accordingly, in all
subsequent algorithms we take \(\pi: V_G \to V_H \cup \{\perp\}\),
with \(\pi(u) = \perp\) meaning that \(u\) is left unmatched in
\(V_H\).

\section{Approximation Algorithm for List-Restricted MaxQAP} \label{sec:list-restricted}

For each $u \in V_G$, recall that a list $\mathcal{L}(u)$ is a list of nodes that $u$ can be matched with and $\mathcal{L}(u) \geq n - k$.
We present an $O(\sqrt{n}+k)$-approximation algorithm for this problem. 
Based on the integer programming for MaxQAP by Adams and Johnson \cite{adams1994quadratic}, where $x_{up}$ and $y_{upvq}$ are binary variables that indicate the event \say{$(u,p) \in M$} and \say{$(u,p),(v,q) \in M$}, we add the additional conditions that prevent each node $u$ to be matched to nodes that are not in $\mathcal{L}(u)$, i.e., $x_{up} =0 \text{ for all } u \in V_G \text{ and } p \in V_H\backslash\mathcal{L}(u)$. Our modified relaxed LP is then stated as follows:\\

\vspace{-0.1cm}
\noindent\hspace{-0.05cm}$
\begin{array}{lll@{}ll}
    \textbf{(Relaxed LP1)}\hspace*{0.1cm}&\text{Maximize}& \displaystyle \sum_{u,v \in V_G} \sum_{p,q \in V_H} w_G(&u,v)w_H(p,q)y_{upvq}\\
    &\text{subject to}& \displaystyle\sum_{p \in V_H} x_{up} \leq 1 & \text{for all } u \in V_G;&\\
    && \displaystyle\sum_{u \in V_G} x_{up} \leq 1 & \text{for all } p \in V_H;&\\
    && \displaystyle\sum_{p \in V_H} y_{upvq} \leq x_{vq} & \text{for all } u,v \in V_G \text{ and } q \in V_H;&\\
    && \displaystyle\sum_{u \in V_G} y_{upvq} \leq x_{vq} & \text{for all } v \in V_G \text{ and } p,q \in V_H;&\\
    && y_{upvq} = y_{vqup} & \text{for all } u,v \in V_G \text{ and } p,q \in V_H;&\\
    && x_{up} \in [0,1] & \text{for all } u \in V_G \text{ and } p \in \mathcal{L}(u);&\\
    && x_{up} =0 & \text{for all } u \in V_G \text{ and } p \in V_H\backslash\mathcal{L}(u);&\\
    && y_{upvq} \in [0,1] & \text{for all } u,v \in V_G \text{ and } p,q \in V_H.&
\end{array}$\\

Our algorithm for List-Restricted MaxQAP is as follows:

\vspace{0.1cm}
\noindent \textbf{Algorithm 1:}
\vspace{-0.1cm}
\begin{center}
\begin{adjustwidth}{0cm}{}
\begin{tabular}{p{1.4cm} p{10.6cm}}
    \textbf{Input:} & $G = (V_G,E_G,w_G)$ and $H = (V_H,E_H,w_H)$ such that $|V_G| = |V_H| = n$, \\
    & and a nonempty set $\mathcal{L}(u) \subseteq V_H$ for each $u \in V_G$\\
    \textbf{Output:} & A compatible matching $M$ between $V_G$ and $V_H$
\end{tabular}
\end{adjustwidth}
\end{center}
\vspace{-0.1cm}
\begin{compactenum}
    \item Solve Relaxed LP1 and obtain an optimal solution: $(x^{*}_{up})_{u \in V_G, p \in V_H}$ and $(y^{*}_{upvq})_{u,v \in V_G, p,q \in V_H}$. 
    \item Partition $V_G$ randomly into $G_L$ and $G_R$, such that $|G_L| = \lceil n/2 \rceil$ and $G_R = V_G \backslash G_L$. Do the same for $V_H$ to obtain $H_L$ and $H_R$.
    \item For each $p \in H_L$, let $p$ choose a node from $G_L \cup \{ \perp \}$ with probability $x^*_{up}$ for $u \in G_L$ and $1-\sum_{u \in G_L}x^*_{up}$ for $\perp$. Then, define a function $\pi : H_L \to G_L \cup \{\perp\}$ such that $\pi(p)$ is the node that $p$ have chosen.
    \item For each $u \in G_L$, let $\pi^{-1}(u)$ be a set of node $p \in H_L$ such that $\pi(p) = u$. For each $u \in G_L$ such that $\pi^{-1}(u) \neq \emptyset$, randomly choose an element $p$ in $\pi^{-1}(u)$ and assign $(u,p)$ to a compatible matching $M_L$. 
    \item For each $v \in G_R$ and $q \in H_R$, let $w(v,q) = \sum_{(u,p) \in M_L} w_G(u,v)w_H(p,q)$ if $q \in \mathcal{L}(v)$, and $w(v,q) = 0$ otherwise. Let $M_R$ be a maximum compatible matching between $G_R$ and $H_R$ with respect to $w$ that does not contain zero-weighted edge. The compatible matching $M_R$ can be calculated in polynomial time using the maximum matching algorithm.
    \item Return $M = M_L \cup M_R$.\\
\end{compactenum}

To show that Algorithm 1 is indeed an $O(\sqrt{n}+k)$-approximation algorithm, we let $LP^*$ be the optimal value given by an optimal solution of Relaxed LP1. For each $p \in V_H$, let $\mathcal{W}_p$ be the set of $\lceil (\sqrt{n+k^2}+k)/2 \rceil$ nodes in $V_H$ whose weights of the edges from these nodes to $p$ are the largest. Then, we have
\begin{align*}
    LP^* &= \sum_{u,v \in V_G} \sum_{p,q \in V_H} w_G(u,v)w_H(p,q)y^*_{upvq}\\
    &=\sum_{u,v \in V_G} \sum_{\substack{p \in V_H \\ q \in \mathcal{W}_p}} \hspace*{-0.05cm} w_G(u,v)w_H(p,q)y^*_{upvq} + \sum_{u,v \in V_G} \hspace*{-0.15cm} \sum_{\substack{p \in V_H \\ q \in V_H \backslash \mathcal{W}_p}} \hspace*{-0.3cm} w_G(u,v)w_H(p,q)y^*_{upvq}.
\end{align*}
\noindent Denote the first sum as $LP^*_1$ and the second sum as $LP^*_2$. For any compatible matching $M$ between $V_G$ and $V_H$, let $Obj(M) = \sum\limits_{(u,p),(v,q) \in M} w_G(u,v)w_H(p,q).$ However, in this analysis of the algorithm which provides two compatible matchings, $M_L$ and $M_R$, between each pair of the corresponding partites of $V_G$ and $V_H$, we mainly focus on $Obj'(M_L,M_R) = \sum_{\substack{(u,p) \in M_L \\ (v,q) \in M_R}} w_G(u,v)w_H(p,q)$ and utilize the fact that $Obj(M_L \cup M_R) \geq Obj'(M_L,M_R)$ in our analysis. 
For a fixed $M_L$, by the choice of constructing $M_R$, we have $Obj'(M_L,M_R) \geq Obj'(M_L,M'_R)$ for any other compatible matching $M'_R$ between $G_R$ and $H_R$.  In the next lemma, we set \(M'_R := M''_{\mathrm{rand}}\), a random matching
between \(G_R\) and \(H_R\). The formal construction of \(M''_{\mathrm{rand}}\)
and the lemma's proof are deferred to Appendix~\ref{appendix:lemma1}.

\begin{lemma} \label{lem:alg1.1}
    $\EX[Obj'(M_L,M_R)] \geq LP^*_2/O(\sqrt{n}+k)$.
\end{lemma}

\paragraph{Proof sketch.}
Randomly split $V_G,V_H$ into $G_L,H_L,G_R,H_R$ as in Algorithm 1. By Lemma~\ref{lem:rand},
\[
\Pr\big[(u,p)\!\in\! M_L \wedge (v,q)\!\in\! M'_{\mathrm{rand}} \mid u\!\in\! G_L,p\!\in\! H_L,v\!\in\! G_R,q\!\in\! H_R\big]
\ge \tfrac{x^*_{up}}{2|G_R|}\ge \tfrac{x^*_{up}}{n}.
\]
Since $\Pr[u\!\in\! G_L,p\!\in\! H_L,v\!\in\! G_R,q\!\in\! H_R]\ge 1/32$ for $n\!\ge\!2$ (Lemma \ref{lem:withoutrandomG_L}),
$\Pr\big[(u,p)\!\in\! M_L \wedge (v,q)\!\in\! M'_{\mathrm{rand}}\big]\ge \tfrac{x^*_{up}}{32n}$.
As $M_L$ only use edges with $x^*_{up}\!>\!0$ (i.e., $p\!\in\!\mathcal{L}(u)$), they are compatible. Consider compatible  $M''_{\mathrm{rand}}=M'_{\mathrm{rand}}\cap E_{GH}$, yielding
$\Pr\big[(u,p)\!\in\! M_L \wedge (v,q)\!\in\! M''_{\mathrm{rand}}\big]
\ge \tfrac{x^*_{up}}{32n}\quad \text{if } q\!\in\!\mathcal{L}(v)$,
and $0$ otherwise. Taking expectations,
\[
\mathbb{E}[Obj'(M_L, M''_{\rm rand})]\ge \frac{1}{32n}\sum_{u,v} w_G(u,v) \sum_{p} x^*_{up} \sum_{q\in \mathcal{L}(v)} w_H(p,q).
\]
Restricting to $q\in \mathcal{L}(v)\cap\mathcal{W}_p$ and using
$|\mathcal{L}(v)\cap\mathcal{W}_p|\ge (\sqrt{n+k^2}-k)/2$ gives
\small \[
\mathbb{E}[Obj'(M_L, M''_{\rm rand})]\ge \frac{1}{32n}\sum_{u,v} w_G(u,v)\sum_{p} x^*_{up}\Big[\frac{\sqrt{n+k^2}-k}{2}\cdot \min_{q\in \mathcal{L}(v)\cap\mathcal{W}_p} w_H(p,q)\Big].
\]
\normalsize
Meanwhile, since $\sum_{q} y^*_{upvq}\le x^*_{up}$, the quantity $LP^*_2$ 
(which places mass on
$q\in V_H\setminus \mathcal{W}_p$) is at most
$\sum_{u,v} w_G(u,v)\sum_{p} x^*_{up}\Big[\min_{q\in \mathcal{L}(v)\cap \mathcal{W}_p} w_H(p,q)\Big]$.
Hence,
$\mathbb{E}[Obj'(M_L,M_R)]\ge \mathbb{E}[Obj'(M_L, M''_{\rm rand})] \ge \frac{(\sqrt{n+k^2}-k)LP^*_2}{64n} = \frac{LP^*_2}{O(\sqrt{n}+k)}$.
\qed

Whereas Lemma~\ref{lem:alg1.1} uses \(M'_{\mathrm{rand}}\) to bound \(LP_2^*\),
our analysis of \(LP_1^*\) employs a different auxiliary matching,
\(M'_{\mathrm{star}}\), which leverages the set $\mathcal{S}^p_R$ of each node $p$ defined as follows. The details for this construction and the properties of the matching can be found in Appendix \ref{appendix:Lemma2}.

\paragraph{Definition of $M'_{\rm star}$.}
Define \(l: V_H \to V_H\) by, for each \(q \in V_H\), \\
  $l(q) \in \arg\max_{p \in V_H} \; \sum_{u,v \in V_G} w_G(u,v)\, w_H(p,q)\, y^{*}_{upvq}$.
For \(p \in V_H\), set
  $l^{-1}(p) \coloneqq \{\, q \in V_H : l(q) = p \,\}$.
Given subsets \(H_L, H_R \subseteq V_H\), define for each \(p \in H_L\),
  $\mathcal{S}_R^{p} \coloneqq l^{-1}(p) \cap H_R$.
We notice that for any distinct \(p,p' \in H_L\),
  $\mathcal{S}_R^{p} \cap \mathcal{S}_R^{p'} = \emptyset$.

Given $\mathcal{S}_R^{p}$ ($p\in H_L$), define for each $p$ the matrix
$Z_p=[z^p_{vq}]_{v\in G_R,\,q\in \mathcal{S}_R^{p}}$ by
$z^p_{vq}=y_{\pi(p)pvq}/x_{\pi(p)p}$ if $\pi(p)\neq\perp$ and
$z^p_{vq}=0$ otherwise. Each $Z_p$ is a fractional matching on
$G_R\times \mathcal{S}_R^{p}$ (row/column sums $\le 1$), hence admits a
convex decomposition
$
Z_p \;=\; \sum_{M\in\mathcal{M}_p} \alpha_M \, V_M,$
where $\mathcal{M}_p$ is the set of integral matchings on
$G_R\times \mathcal{S}_R^{p}$ and, for each $M\in\mathcal{M}_p$, the
matrix $V_M\in\{0,1\}^{G_R\times \mathcal{S}_R^{p}}$ (indexed by rows
$v\in G_R$ and columns $q\in \mathcal{S}_R^{p}$) is the incidence
(biadjacency) matrix of $M$, defined entrywise by $(V_M)_{vq} = 1$ if $(v,q)\in M$ and $(V_M)_{vq} = 0$ otherwise.

Sample one $M\in\mathcal{M}_p$ with probability $\alpha_M$ and define
$\tau_p:\mathcal{S}_R^{p}\to G_R\cup\{\perp\}$ by $\tau_p(q)=v$ if
$(v,q)\in M$ and $\tau_p(q)=\perp$ otherwise. Since the
$\mathcal{S}_R^{p}$’s are disjoint, the domains $\operatorname{dom}(\tau_p)$ are
disjoint, so we may set $\tau:H_R\to G_R\cup\{\perp\}$ by
$\tau(q)=\tau_p(q)$ when $q\in \mathcal{S}_R^{p}$, and $\tau(q)=\perp$
otherwise. Finally, for each $v\in G_R$ with $\tau^{-1}(v)\neq\emptyset$,
choose one $q\in\tau^{-1}(v)$ uniformly at random and include $(v,q)$ in
$M'_{\text{star}}$. The resulting set of edges $M'_{\text{star}}$ is a matching on
$G_R\times H_R$.

We obtain the following lemma, of which the full proof can be found in Appendix \ref{appendix:Lemma2}, from the construction of $M'_{\rm star}$. 

\begin{lemma} \label{lem:alg1.2}
    $\EX[Obj'(M_L,M_R)] \geq LP^*_1/O(\sqrt{n}+k) $.
\end{lemma}

\paragraph{Proof sketch.}
From the LP, $y^*_{upvq}=0$
whenever $q\notin\mathcal{L}(v)$. Hence, in the convex decomposition
$Z_p=\sum_{M\in\mathcal{M}_p}\alpha_M V_M$, any matching $M$ containing a
non-edge $(v,q)\notin E_{GH}$ must have $\alpha_M=0$, so the sampled
$M'_{\mathrm{star}}$ is compatible. (By construction, $M_L$ is also compatible.)
Taking expectations with the probability given by Lemma \ref{lem:star} and Lemma \ref{lem:withoutrandomG_L} provides
$\EX[Obj'(M_L,M'_{\mathrm{star}})]
\ge \frac{1}{128}\sum\limits_{q\in V_H}\sum\limits_{u,v\in V_G}
y^*_{u l(q) v q}\,w_G(u,v)w_H(l(q),q)$.
By definition of $l(q)$ as a maximizer over $p$, this is at least the
average over $p\in\mathcal{W}_q$:
\[
\EX[Obj'(M_L,M'_{\mathrm{star}})]\;\ge\;
\frac{1}{128}\sum_{q}\frac{1}{|\mathcal{W}_q|}
\sum_{p\in\mathcal{W}_q}\sum_{u,v} y^*_{upvq} w_G(u,v)w_H(p,q).
\]
Finally, the fact that $|\mathcal{W}_q|\le (\sqrt{n+k^2}+k+2)/2$ gives
$\EX[Obj'(M_L,M'_R)] \geq \EX[Obj'(M_L,M'_{\mathrm{star}})] \;\ge\; \frac{LP_1^*}{64(\sqrt{n+k^2}+k+2)} = \frac{LP_1^*}{O(\sqrt{n}+k)}$. 
 \qed
We are now ready to demonstrate the main theorem of this section.

\begin{theorem}
    Algorithm 1 is an $O(\sqrt{n}+k)$-approximation algorithm for List-Restricted MaxQAP.
\end{theorem}
\begin{proof}
    Let $M^*$ be an optimal solution of List-Restricted MaxQAP and $M = M_L \cup M_R$ be a matching given by Algorithm 1. Following Lemma \ref{lem:alg1.1} and Lemma \ref{lem:alg1.2}, we have 
    \[\EX[Obj(M)] \geq \EX[Obj'(M_L,M_R)] \geq  \frac{LP^*_1+LP^*_2}{O(\sqrt{n}+k)}  = \frac{LP^*}{O(\sqrt{n}+k)}  \geq \frac{Obj(M^*)}{O(\sqrt{n}+k)}.\] \qed
\end{proof}

\section{Approximation Algorithm for MaxQbAP}
\label{sec:QbAP}

In this section, we present an approximation algorithm for dup-MaxQbAP.
We again build a linear program on the MaxQAP integer formulation of Adams and Johnson~\cite{adams1994quadratic}.
Because each node may be paired with up to \(b\) nodes on the opposite side, the capacity constraint changes from \(\sum_{u\in V_G} x_{up}=1\) to
$\sum_{u\in V_G} x_{up}=b$.
For the same reason, we also impose
$\sum_{p\in V_H} y_{upvq}= b\,x_{vq}$.
\\

\vspace{-0.1cm}
\noindent\hspace{-0.05cm}$
\begin{array}{lll@{}ll}
    \textbf{(Relaxed LP2)}\hspace*{0.1cm}&\text{Maximize}& \displaystyle \sum_{u,v \in V_G} \sum_{p,q \in V_H} w_G(u,&v)w_H(p,q)y_{upvq}\\
    &\text{subject to}& \displaystyle\sum_{p \in V_H} x_{up} \leq b & \text{for all } u \in V_G;&\\
    && \displaystyle\sum_{u \in V_G} x_{up} \leq b & \text{for all } p \in V_H;&\\
    && \displaystyle\sum_{p \in V_H} y_{upvq} \leq bx_{vq} & \text{for all } u,v \in V_G \text{ and } q \in V_H;&\\
    && \displaystyle\sum_{u \in V_G} y_{upvq} \leq bx_{vq} & \text{for all } v \in V_G \text{ and } p,q \in V_H;&\\
    && y_{upvq} = y_{vqup} & \text{for all } u,v \in V_G \text{ and } p,q \in V_H;&\\
    && x_{up} \in [0,1] & \text{for all } u \in V_G \text{ and } p \in V_H;&\\
    && y_{upvq} \in [0,1] & \text{for all } u,v \in V_G \text{ and } p,q \in V_H.&
\end{array}$\\

Our algorithm for dup-MaxQbAP is as follows:

\vspace{0.1cm}
\noindent \textbf{Algorithm 2:}
\vspace{-0.1cm}
\begin{center}
\begin{adjustwidth}{0cm}{}
\begin{tabular}{p{1.4cm} p{10.6cm}}
    \textbf{Input:} & $G = (V_G,E_G,w_G)$ and $H = (V_H,E_H,w_H)$ such that $|V_G| = |V_H| = n$\\
    \textbf{Output:} & A $b$-matching $\leftindex^b{M}$ between $V_G$ and $V_H$
\end{tabular}
\end{adjustwidth}
\end{center}
\vspace{-0.1cm}
\begin{compactenum}
    \item Solve Relaxed LP2 and obtain an optimal solution: $(x^{*}_{up})_{u \in V_G, p \in V_H}$ and $(y^{*}_{upvq})_{u,v \in V_G, p,q \in V_H}$. 
    \item Partition $V_G$ randomly into $G_L$ and $G_R$, such that $|G_L| = \lceil n/2 \rceil$ and $G_R = V_G \backslash G_L$. Do the same for $V_H$ to obtain $H_L$ and $H_R$.
    \item For each $i = 1,2,...,b$:
    \begin{enumerate}
        \item For each $p \in H_L$, let $p$ choose a node from $G_L \cup \{ \perp \}$ with probability $x^*_{up}/b$ for $u \in G_L$ and $1-\sum_{u \in G_L}x^*_{up}/b$ for $\perp$. Then, define a function $\pi_i : H_L \to G_L \cup \{\perp\}$ such that $\pi_i(p)$ is the node that $p$ have chosen.
        \item For each $u \in G_L$, let $\pi_i^{-1}(u)$ be a set of node $p \in H_L$ such that $\pi_i(p) = u$. For each $u \in G_L$ such that $\pi_i^{-1}(u) \neq \emptyset$, randomly choose an element $p$ in $\pi_i^{-1}(u)$ and assign $(u,p)$ to a matching $M_L^{(i)}$. Then add edges in $E_{GH}$ between unmatched nodes in $G_L$ and $H_L$ to the matching $M_L^{(i)}$ at random, so that $M_L^{(i)}$ is a perfect matching between $G_L$ and $H_L$.
    \end{enumerate}
    \item Let $\leftindex^b{M}_L = \bigcup_{i=1}^b M_L^{(i)}$.
    \item For each $v \in G_R$ and $q \in H_R$, let $w(v,q) = \sum_{(u,p) \in \leftindex^b{M}_L} w_G(u,v)w_H(p,q)$. Let $\leftindex^b{M}_R$ be a maximum $b$-matching between $G_R$ and $H_R$ with respect to $w$. This $b$-matching can be obtained in polynomial time by using an algorithm for the maximum-weight $b$-matching problem (the maximum weight upper degree-constrainted subgraph with unit capacity problem) proposed by Gabow~\cite{10.1145/800061.808776}.
    \item Return $\leftindex^b{M} = \leftindex^b{M}_L \cup \leftindex^b{M}_R$. \\
\end{compactenum}

The proof can be done similarly to the proof of Algorithm 1. Let $LP^*$ be an optimal value of Relaxed LP2 above. For each $p \in V_H$, let $\mathcal{U}_p$ be the set of $\lceil \sqrt{n}/\sqrt{b} \rceil$ nodes in $V_H$ whose weights of the edges from these nodes to $p$ are the largest. Then we have
\begin{align*}
    LP^* &= \sum_{u,v \in V_G} \sum_{p,q \in V_H} w_G(u,v)w_H(p,q)y^*_{upvq}\\
    &=\sum_{u,v \in V_G} \sum_{\substack{p \in V_H \\ q \in \mathcal{U}_p}} \hspace*{-0.05cm} w_G(u,v)w_H(p,q)y^*_{upvq} + \sum_{u,v \in V_G} \hspace*{-0.15cm} \sum_{\substack{p \in V_H \\ q \in V_H \backslash \mathcal{U}_p}} \hspace*{-0.3cm} w_G(u,v)w_H(p,q)y^*_{upvq}.
\end{align*}
\noindent Denote the first sum as $LP^*_1$ and the second sum as $LP^*_2$. 
Recall that \( {}^b\!M = {}^b\!M_L \cup {}^b\!M_R \).
Define
$
Obj'\!\left({}^b\!M_L, {}^b\!M_R\right)
= \sum_{\substack{(u,p)\in {}^b\!M_L \\ (v,q)\in {}^b\!M_R}}
w_G(u,v)\, w_H(p,q),
$
and evaluate Algorithm~2 using \(Obj'\) rather than the conventional
\(Obj({}^b\!M)\), as in the analysis of Algorithm~1.
We then compare \(LP_2^*\) with the value achieved by Algorithm~2 via an argument
parallel to the proof of Lemma~\ref{lem:alg1.1}, yielding the following lemma. We give the full proof the lemma in Appendix \ref{appendix:lemma3}.

\begin{lemma} \label{lem:alg4.1}
$\EX[Obj'(\leftindex^b{M}_L,\leftindex^b{M}_R)] \geq LP^*_2/O(\sqrt{bn})$.
\end{lemma}

\paragraph{Proof sketch.}
 Consider $b$ independent random matchings $M'^{(1)}_{\rm rand},\dots,M'^{(b)}_{\rm rand}$ between $G_R$ and $H_R$, and define
${}^{b}\!M'_{\rm rand} \coloneqq \bigcup_{i=1}^b M'^{(i)}_{\rm rand}$.

By Lemma~\ref{lem:rand}, in each round $i$ we have
$\Pr\bigl[(u,p)\in M_L^{(i)} \,\big|\, u\in G_L,\;p\in H_L\bigr] \;\ge\; \frac{x^*_{up}}{2b}$, and
$\Pr\bigl[(v,q)\in M_{\rm rand}'^{(i)} \,\big|\, v\in G_R,\;q\in H_R\bigr] \;=\; \frac{1}{|G_R|} \;\ge\; \frac{2}{n}$.
Since the $2b$ events $\{(u,p)\in M_L^{(i)}\}$ and $\{(v,q)\in M_{\rm rand}'^{(i)}\}$ are pairwise independent,
$\Pr\bigl[(u,p)\in {}^{b}\!M_L \text{ and } (v,q)\in {}^{b}\!M'_{\rm rand}\bigr]
= \bigl(1-\Pr[(u,p)\notin {}^{b}\!M_L]\bigr)\bigl(1-\Pr[(v,q)\notin {}^{b}\!M'_{\rm rand}]\bigr)$. Since
$
\Pr[(u,p)\notin {}^{b}\!M_L] \le \Bigl(1-\frac{x^*_{up}}{2b}\Bigr)^b, \;
\Pr[(v,q)\notin {}^{b}\!M'_{\rm rand}] \le \Bigl(1-\frac{2}{n}\Bigr)^b,
$
we apply $1+x\le e^x$ and $1-e^{-x} \ge (1-1/e)x$ (for $0\le x\le 1$) to obtain
\[
\Pr\bigl[(u,p)\in {}^{b}\!M_L \text{ and } (v,q)\in {}^{b}\!M'_{\rm rand}\bigr]
\;\ge\; \Bigl(1-\frac{1}{e}\Bigr)^2 \cdot \frac{b x^*_{up}}{n}.
\]

Lemma~\ref{lem:withoutrandomG_L} shows that the additional randomness from the partition into $G_L,G_R,H_L,H_R$ only incurs a further constant-factor loss, so there exists
$c_1 \;=\; \frac{32}{(1-1/e)^2}$
such that
$\Pr\bigl[(u,p)\in {}^{b}\!M_L \text{ and } (v,q)\in {}^{b}\!M'_{\rm rand}\bigr]
\;\ge\; \frac{b x^*_{up}}{c_1 n}$
for all $u,v,p,q$.

By linearity of expectation, this probability bound yields a lower bound on
$\EX\bigl[Obj'({}^{b}\!M_L,{}^{b}\!M'_{\rm rand})\bigr]$
in terms of the LP solution. Using the fact that $|\mathcal{U}_p|\ge \sqrt{n}/\sqrt{b}$ and the constraint $\sum_{q\in V_H} y^*_{upvq} \le b x^*_{up}$, one can extract a factor $\sqrt{n}/\sqrt{b}$ and show
\[
\EX\bigl[Obj'({}^{b}\!M_L,{}^{b}\!M'_{\rm rand})\bigr]
\ge
\frac{1}{c_1\sqrt{bn}}
\sum_{u,v\in V_G}
\sum_{\substack{p\in V_H \\ q\in V_H\setminus \mathcal{U}_p}}
w_G(u,v)w_H(p,q)\,y^*_{upvq}
=
\frac{LP_2^*}{c_1\sqrt{bn}}.
\]

Finally, we obtain that
$\EX\bigl[Obj'({}^{b}\!M_L,{}^{b}\!M_R)\bigr]
\;\ge\;
\EX\bigl[Obj'({}^{b}\!M_L,{}^{b}\!M'_{\rm rand})\bigr]
\;\ge\;
\frac{LP_2^*}{c_1\sqrt{bn}}
\;=\;
\frac{LP_2^*}{O(\sqrt{bn})}.
$ \qed

In the next lemma, we discuss the relationship between our objective values and $LP^*_1$. We give the full proof the lemma in Appendix \ref{appendix:lemma4}.

\begin{lemma} \label{lem:alg4.2}
    $\EX[Obj'(\leftindex^b{M}_L,\leftindex^b{M}_R)] \geq LP^*_1/O(\sqrt{bn})$.
\end{lemma}

\paragraph{Proof sketch.}
We scale
$x_{up} \coloneqq \frac{x^*_{up}}{b}$, 
$y_{upvq} \coloneqq \frac{y^*_{upvq}}{b^2}$,
and run Algorithm~C for $b$ independent iterations using these scaled values. Let $M_{\rm star}'^{(i)}$ be the matching produced in round $i$, and set
${}^{b}\!M'_{\rm star} \coloneqq \bigcup_{i=1}^b M_{\rm star}'^{(i)}$.

By Lemma~\ref{lem:star}, conditioned on $u\in G_L$, $p\in H_L$, $v\in G_R$ and $q\in \mathcal{S}_R^p$ (with $\mathcal{S}_R^{p} = l^{-1}(p)\cap H_R$), each iteration satisfies
$
\Pr\bigl[(u,p)\in M_L'^{(i)} \text{ and } (v,q)\in M_{\rm star}'^{(i)}\bigr]
\;\ge\; \frac{y^*_{upvq}}{4b^2}.
$
The fact that $b$ iterations are independent gives 
$
\Pr\bigl[(u,p)\in {}^{b}\!M'_L \text{ and } (v,q)\in {}^{b}\!M'_{\rm star}\bigr]
\;\ge\;
1 - \Bigl(1-\frac{y^*_{upvq}}{4b^2}\Bigr)^b.
$
Applying the standard bound $1-e^{-x} \ge (1-1/e)x$ for $0\le x\le 1$ yields
\[
\Pr\bigl[(u,p)\in {}^{b}\!M'_L \text{ and } (v,q)\in {}^{b}\!M'_{\rm star}\bigr]
\;\ge\;
\Bigl(1-\frac{1}{e}\Bigr)\cdot \frac{y^*_{upvq}}{4b}.
\]

Hence, there exists a constant
$c_2 \coloneqq \frac{128}{1-1/e}$
such that, for any $u,v\in V_G$ and $p,q\in V_H$,
$\Pr\!\bigl[(u,p)\in {}^{b}\!M_L \text{ and } (v,q)\in {}^{b}\!M'_{\rm star} \,\bigm|\, p=l(q)\bigr]
\;\ge\;
\frac{y^*_{upvq}}{c_2\,b}$ and 
$\Pr\bigl[(u,p)\in {}^{b}\!M'_L \text{ and } (v,q)\in {}^{b}\!M'_{\rm star} \,\bigm|\, p\neq l(q)\bigr]=0$.

Using the same averaging argument as in the proof of Lemma~\ref{lem:alg1.2}, together with the fact that $1/|\mathcal{U}_q| \ge \sqrt{b}/(2\sqrt{n})$, we obtain
\[
\EX\bigl[Obj'({}^{b}\!M'_L,{}^{b}\!M'_{\rm star})\bigr]
\ge
\frac{1}{O(\sqrt{bn})}
\sum_{u,v\in V_G}
\sum_{\substack{q\in V_H \\ p\in \mathcal{U}_q}}
w_G(u,v)w_H(p,q)\,y^*_{upvq}
=
\frac{LP_1^*}{O(\sqrt{bn})}.
\]
Therefore, 
$\EX\bigl[Obj'({}^{b}\!M_L,{}^{b}\!M_R)\bigr]
\;\ge\;
\EX\bigl[Obj'({}^{b}\!M'_L,{}^{b}\!M'_{\rm star})\bigr]
\;\ge\;
\frac{LP_1^*}{O(\sqrt{bn})}$. \qed

Finally, we conclude the approximation ratio of Algorithm 2 as follow. 

\begin{theorem}
    Algorithm 2 is an $O(\sqrt{bn})$-approximation algorithm for dup-Max\allowbreak QbAP, and hence, an $O(\sqrt{bn})$-approximation algorithm for MaxQbAP.
\end{theorem}

\begin{proof}
    Let $\leftindex^b{M}^*$ be an optimal solution of MaxQbAP and $\leftindex^b{M} = \leftindex^b{M}_L \cup \leftindex^b{M}_R$ be a matching given by Algorithm 2. Following Lemma~\ref{lem:alg4.1} and Lemma~\ref{lem:alg4.2}, we have 
    \[\EX[Obj(\leftindex^b{M})] \geq \EX[Obj'(\leftindex^b{M}_L,\leftindex^b{M}_R)] \geq \frac{LP^*_1 + LP^*_2}{O(\sqrt{bn})} = \frac{LP^*}{O(\sqrt{bn})} \geq \frac{Obj(\leftindex^b{M}^*)}{O(\sqrt{bn})}.\] \qed
\end{proof}

\section{Conclusion and Future Works}

While MaxQAP has been extensively studied due to its wide range of applications, its variants have received significantly less attention in the literature. In particular, to the best of our knowledge, no prior work has addressed approximation algorithms for any variant of MaxQAP. To initiate progress in this direction, we investigate two natural variants: MaxQAP with list restrictions and the Maximum Quadratic \( b \)-Matching Problem (MaxQbAP). These two natural variants have also been studied in problems related to MaxQAP, such as graph isomorphism and MEEBM.

Using the non-standard randomized rounding technique and the intricate approximation-ratio analysis of \cite{makarychev2014maximum}, we develop approximation algorithms for both variants. Our approximation ratio asymptotically matches the best-known ratio for MaxQAP when each node’s restricted list has size \( n - O(\sqrt{n}) \) and \( b \in O(1) \), suggesting that improving this ratio may require fundamentally new techniques.

In the future, we will implement the algorithms to gain computational results as validation of effectiveness and runtime, and possibly to indicate windows of the ratio's improvement, as well as offer efficient methods to speed up the execution.
Also, we will improve the approximation ratio for the List-Restricted MaxQAP with lists of arbitrary size, especially for large $k$, through a new approach of proving or new approximation algorithms specific to the cases.

\bibliographystyle{plain} 
\bibliography{sample}

\newpage
\begin{subappendices}
\renewcommand{\thesection}{\Alph{section}}
\setcounter{lemma}{0}
\renewcommand{\thelemma}{\Alph{section}\arabic{lemma}}
\section{Relationship between MaxQbAP and dup-MaxQbAP} \label{appendix:dup-maxqbap}

Since MaxQbAP differs from dup-MaxQbAP only in the objective function, we can use the same approximation algorithm on both problems with the only difference in the approximation ratios, as stated in the following lemma. 
\begin{lemma} \label{lem:dup}
    An $\alpha$-approximation algorithm for dup-MaxQbAP is a $2\alpha$-approxi-mation algorithm for MaxQbAP.
\end{lemma}
\begin{proof}
    Let $Obj$ and $Obj_{dup}$ denote the objective functions of MaxQbAP and dup-MaxQbAP, respectively. Given an instance of MaxQbAP, we first observe that in dup-MaxQbAP, each pair of edge weights $w_G(u,v)w_H(p,q)$ for $\{u,v\} \in E_G$ and $\{p,q\} \in E_H$ can appear in the objective function at most twice; one when $(u,p),(v,q) \in {}^b\!M$ and the other when $(u,q),(v,p) \in {}^b\!M$. Furthermore, for any $b$-matching $^b\!M$, we know that the pair $w_G(u,v)w_H(p,q)$ is in the objective function of MaxQbAP for $^b\!M$ if and only if it is in the objective function of dup-MaxQbAP (can be once or twice). Thus, we have \[Obj_{dup}(^b\!M) \geq Obj(^b\!M) \geq \frac{Obj_{dup}(^b\!M)}{2}\] for any $b$-matching $^b\!M$ between $V_G$ and $V_H$. Let $Opt$ be the optimal value given by an optimal solution $^b\!M^*$ of MaxQbAP, and let $Opt_{dup}$ be the optimal value of dup-MaxQbAP on this instance. Assume that we have an $\alpha$-approximation algorithm for dup-MaxQbAP. Then we can find a $b$-matching $^b\!M_{Sol}$ such that $Obj_{dup}(^b\!M_{Sol}) \geq Opt_{dup}/\alpha$. Hence, we obtain the following:
    \[Obj(^b\!M_{Sol}) \geq \frac{Obj_{dup}(^b\!M_{Sol})}{2} \geq \frac{Opt_{dup}}{2\alpha} \geq \frac{Obj_{dup}(^b\!M^*)}{2\alpha} \geq \frac{Obj(^b\!M^*)}{2\alpha} = \frac{Opt}{2\alpha}.\] \qed
\end{proof}

\section{Proof of Lemma \ref{lem:alg1.1}} \label{appendix:lemma1}
\setcounter{lemma}{0}
We first give a lemma which we have used for proving Lemma \ref{lem:alg1.1}.

\begin{lemma}[Lemma 5.1 of \cite{nguyen2021maximum}] \label{lem:ind1}
    Let $S$ be a random subset of a set $V$ and let $v \in V$. Assume that for all $s,s' \in V$ such that $s \neq s'$, the event that $s \in S$ is independent to the event that $s' \in S$. If we pick an element from $S$ at random, the probability of having $v$ as the chosen element is at least $\Pr[v \in S]/(\EX[|S|]+1)$.
\end{lemma}

Then, we formally define how we construct the matching $M'_{\rm rand}$ in the following algorithm.

\vspace{0.1cm}
\noindent \textbf{Algorithm B:}
\vspace{-0.1cm}
\begin{center}
\begin{adjustwidth}{0cm}{}
\begin{tabular}{p{1.4cm} p{10.6cm}}
    \textbf{Input:} & Pairwise disjoint sets of nodes $G_L,G_R,H_L,H_R$ such that  $|G_L|=|H_L|$ and $|G_R|=|H_R|$, $x_{up} \in [0,1]$ for each $u \in G_L$ and $p \in H_L $ such that \\
    & \quad $\displaystyle\sum_{u \in G_L}x_{up} \leq 1$ for all $p \in H_L$ and $\displaystyle\sum_{p \in H_L}x_{up} \leq 1$ for all $u \in G_L$\\
     \textbf{Output:} & Two matchings: $M_L$ between $G_L$ and $H_R$ and $M'_{\rm rand}$ between $G_R$ and $H_R$
\end{tabular}
\end{adjustwidth}
\end{center}
\vspace{-0.1cm}
\begin{compactenum}
    \item For each $p \in H_L$, choose a node from $G_L \cup \{ \perp \}$ with probability $x_{up}$ for $u \in G_L$ and $1-\sum_{u \in G_L}x_{up}$ for $\perp$. Then, define a function $\pi : H_L \to G_L \cup \{\perp\}$ such that $\pi(p)$ is the chosen node for each $p$.
    \item For each $u \in G_L$, let $\pi^{-1}(u)$ be the set of node $p \in H_L$ such that $\pi(p) = u$. For each $u \in G_L$ such that $\pi^{-1}(u) \neq \emptyset$, randomly choose an element $p$ in $\pi^{-1}(u)$ and assign $(u,p)$ to a matching $M_L$.
    \item Construct a random matching $M'_{\rm rand}$ between the set $G_R$ and the set $H_R$.
    \item Return $M_L$ and $M'_{\rm rand}$.\\
\end{compactenum}

\begin{lemma} 
    If we proceed according to Algorithm B with the required input, then for $u \in G_L, p \in H_L, v \in G_R, q \in H_R$, we have $$\Pr[(u,p) \in M_L] \geq x_{up}/2$$ as well as $$\Pr[(v,q) \in M'_{\rm rand}] \geq 1/|G_R|;$$ and consequently,
    $$\Pr[(u,p) \in M_L \text{ and } (v,q) \in M'_{\rm rand}] \geq x_{up}/(2|G_R|).$$ \label{lem:rand}
\end{lemma}
\begin{proof}
    Let $u \in G_L$ and $p \in H_L$. Following the construction of the function $\pi$ and the condition of the input, we have \[\EX[|\pi^{-1}(u)|] = \sum_{p \in H_L}\Pr[\pi(p)=u] = \sum_{p \in H_L}x_{up} \leq 1.\]
    Moreover, for $p,p' \in H_L$ such that $p \neq p'$, the event that $p \in \pi^{-1}(u)$ is independent to the event that $p' \in \pi^{-1}(u)$. Thus, by Lemma \ref{lem:ind1}, we then have \[\Pr[(u,p) \in M_L] \geq \displaystyle\frac{\Pr[p \in \pi^{-1}(u)]}{\EX[|\pi^{-1}(u)|] +1} \geq \frac{\Pr[\pi(p) = u]}{2} = \frac{x_{up}}{2}.\]

    On the other hand, for $v \in G_R$ and $q \in H_R$, we have $\Pr[(v,q) \in M'_{\rm rand}] = 1/|G_R|$ as $M'_{\rm rand}$ is a random matching. Given that $u \in G_L, p \in H_L, v \in G_R, q \in H_R$, since the event that $(u,p) \in M_L$ and the event that $(v,q) \in M'_{\rm rand}$ are independent, we therefore obtain 
\begin{align*}
    \Pr[(u,p) \in M_L \text{ and } (v,q) \in M'_{\rm rand}] &= \Pr[(u,p) \in M_L] \cdot \Pr[(v,q) \in M'_{\rm rand}] \\
    & \geq \frac{x_{up}}{2} \cdot \frac{1}{|G_R|} = \frac{x_{up}}{2|G_R|}.
\end{align*} \qed
\end{proof}

In the above lemma, the probability is considered when the sets $G_L,G_R,H_L, \allowbreak H_R$ are given. However, in both Algorithm 1 and Algorithm 2, these sets are random. Thus, we state the following lemma.

\begin{lemma} \label{lem:withoutrandomG_L}
    Let $G_L,G_R,H_L,H_R$ be randomly chosen as defined in Algorithm 1 and Algorithm 2 and let $A$ be an event of nodes $u,v \in V_G$ and $p,q \in V_H$ that occurs only if $u \in G_L, p \in H_L, v \in G_R, q \in H_R$. Then \[\Pr[A] \geq \frac{\Pr[A | u \in G_L, p \in H_L, v \in G_R, q \in H_R]}{32}.\]
\end{lemma}
\begin{proof}
    Following the definitions of $G_L$, $H_L$, $G_R$, and $H_R$, we have 
    \begin{align*}
        \Pr[u \in G_L, p \in H_L, v \in G_R, q \in H_R] 
        &= \Big(\frac{\lfloor n/2 \rfloor \lceil n/2 \rceil}{n^2} \Big) ^2 \\ &\geq \Big(\frac{(n+1)(n-1)}{4n^2} \Big) ^2 \\ &= \frac{n^4-2n^2+1}{16n^4} \\ & \geq \frac{1}{32}
    \end{align*}
    \noindent when $n \geq 2$. With the condition of the event $A$, we obtain that
    \begin{align*}
        \Pr[A] & = \Pr[A|u \in G_L, p \in H_L, v \in G_R, q \in H_R]\\
        & \qquad \cdot \Pr[u \in G_L, p \in H_L, v \in G_R, q \in H_R]\\
        & \geq \frac{\Pr[A|u \in G_L, p \in H_L, v \in G_R, q \in H_R]}{32}.
    \end{align*} \qed
\end{proof}

We are now ready to demonstrate Lemma \ref{lem:alg1.1}.

\begin{proof} [Proof of Lemma \ref{lem:alg1.1}]
    Given $G_L$, $H_L$, $G_R$, $H_R$ and $x^*_{up}$ from step 1 and 2 of Algorithm 1, we proceed according to Algorithm B using the corresponding sets and $x_{up} = x^*_{up}$ for each $u \in G_L$ and $p \in H_L$. By Lemma \ref{lem:rand}, we have 
    \begin{align*}
        &\Pr[(u,p) \in M_L \text{ and } (v,q) \in M'_{\rm rand}|u \in G_L, p \in H_L, v \in G_R, q \in H_R] \\
        &\qquad \geq \displaystyle \frac{x^*_{up}}{2|G_R|} \geq \frac{x^*_{up}}{2} \cdot \frac{2}{n} = \frac{x^*_{up}}{n}.
    \end{align*}

    \noindent Since the event \say{$(u,p) \in M_L \text{ and } (v,q) \in M'_{\rm rand}$} can occur only if $u \in G_L, p \in H_L, v \in G_R, q \in H_R$, by Lemma \ref{lem:withoutrandomG_L}, we then have \[\Pr[(u,p) \in M_L \text{ and } (v,q) \in M'_{\rm rand}] \geq  \frac{x^*_{up}}{n} \cdot \frac{1}{32} = \frac{x^*_{up}}{32n}.\]

    Note that in the construction of $M_L$, the edge $(u,p)$ when $u \in G_L$ and $p \in H_L$ can be chosen only if $x^*_{up} > 0$, i.e., $p \in \mathcal{L}(u)$. Thus, $M_L$ is compatible. However, for $M'_{\rm rand}$, any edges can be included in the matching. Hence, we consider the matching $M''_{\rm rand} = M'_{\rm rand} \cap E_{GH}$ so that $M''_{\rm rand}$ is compatible. With this definition of $M''_{\rm rand}$, we then have 
    $\Pr[(u,p) \in M_L \text{ and } (v,q) \in M''_{\rm rand}] \geq x^*_{up}/32n$ if $q \in \mathcal{L}(v)$, and equals to $0$ otherwise. Therefore, we get
\begin{align*}
    &\EX[Obj'(M_L,M_R)] \\
    &\qquad \geq \EX[Obj'(M_L,M''_{\rm rand})]\\
    &\qquad = \sum_{u,v \in V_G}\sum_{p,q \in V_H} \Pr[(u,p) \in M_L \text{ and } (v,q) \in M''_{\rm rand}] w_G(u,v)w_H(p,q)\\
    &\qquad = \sum_{u,v \in V_G}\sum_{\substack{p \in V_H \\ q \in \mathcal{L}(v)}} \Pr[(u,p) \in M_L \text{ and } (v,q) \in M''_{\rm rand}] w_G(u,v)w_H(p,q)\\
    &\qquad \geq \sum_{u,v \in V_G}\sum_{\substack{p \in V_H \\ q \in \mathcal{L}(v)}} \frac{x^*_{up}}{32n}w_G(u,v)w_H(p,q)\\
    &\qquad = \frac{1}{32n} \sum_{u,v \in V_G} w_G(u,v) \sum_{p \in V_H} x^*_{up} \sum_{q \in \mathcal{L}(v)} w_H(p,q)\\
    &\qquad \geq \frac{1}{32n} \sum_{u,v \in V_G} w_G(u,v) \sum_{p \in V_H} x^*_{up} \sum_{q \in \mathcal{L}(v) \cap \mathcal{W}_p} w_H(p,q)\\
    &\qquad \geq \frac{1}{32n} \sum_{u,v \in V_G} w_G(u,v) \sum_{p \in V_H} x^*_{up} \bigg[ \frac{\sqrt{n+k^2}-k}{2} \cdot \min_{q \in \mathcal{L}(v) \cap \mathcal{W}_p} w_H(p,q) \bigg]\\
\end{align*}

 The last inequality follows from the fact that $|\mathcal{L}(v) \cap \mathcal{W}_p| = |\mathcal{W}_p \backslash \mathcal{L}(v)^C| \geq |\mathcal{W}_p| - |\mathcal{L}(v)^C| \geq  \dfrac{\sqrt{n+k^2}+k}{2} - k =  \dfrac{\sqrt{n+k^2}-k}{2}$ for all $v \in V_G$ and $p \in V_H$. On the other hand, since $\displaystyle\sum_{q \in V_H} y^*_{upvq} \leq x^*_{up} \text{ for all } u,v \in V_G \text{ and } p \in V_H$, we have
\begin{align*}
    LP^*_2 &= \sum_{u,v \in V_G} \sum_{\substack{p \in V_H \\ q \in V_H \backslash \mathcal{W}_p}} w_G(u,v)w_H(p,q)y^*_{upvq}\\
    &= \sum_{u,v \in V_G} w_G(u,v) \sum_{p \in V_H} \sum_{q \in \mathcal{L}(v) \cap (V_H \backslash \mathcal{W}_p)} w_H(p,q)y^*_{upvq}\\
    &\leq \sum_{u,v \in V_G} w_G(u,v) \sum_{p \in V_H} \sum_{q \in \mathcal{L}(v) \cap (V_H \backslash \mathcal{W}_p)} \bigg[ \max_{q \in \mathcal{L}(v) \cap (V_H \backslash\mathcal{W}_p)} w_H(p,q) \bigg] y^*_{upvq}\\
    &\leq \sum_{u,v \in V_G} w_G(u,v) \sum_{p \in V_H} \bigg[ \max_{q \in \mathcal{L}(v) \cap (V_H \backslash\mathcal{W}_p)} w_H(p,q) \bigg] \sum_{q \in V_H} y^*_{upvq}\\
    &\leq \sum_{u,v \in V_G} w_G(u,v) \sum_{p \in V_H} \bigg[ \max_{q \in \mathcal{L}(v) \cap (V_H \backslash\mathcal{W}_p)} w_H(p,q) \bigg] x^*_{up}\\
    &\leq \sum_{u,v \in V_G} w_G(u,v) \sum_{p \in V_H} \bigg[ \min_{q \in \mathcal{L}(v) \cap \mathcal{W}_p} w_H(p,q) \bigg] x^*_{up}.
\end{align*}

\noindent Therefore, as $\sqrt{n+k^2} + k \leq \sqrt{n}+k+k  = O(\sqrt{n}+k)$, we have $$\EX[Obj'(M_L,M_R)] \geq \dfrac{(\sqrt{n+k^2}-k)LP^*_2}{64n} = \dfrac{LP^*_2}{64(\sqrt{n+k^2}+k)} \geq \dfrac{LP^*_2}{O(\sqrt{n}+k)}.$$ \qed
\end{proof}

\section{Proof of Lemma \ref{lem:alg1.2}} \label{appendix:Lemma2}
\setcounter{lemma}{0}
Recall the function \(l: V_H \to V_H\) defined by, for each \(q \in V_H\),
\begin{equation*}
  l(q) \in \arg\max_{p \in V_H} \; \sum_{u,v \in V_G} w_G(u,v)\, w_H(p,q)\, y^{*}_{upvq}.
\end{equation*} From this function, if the subsets \(H_L, H_R \subseteq V_H\) are given, we define $\mathcal{S}_R^{p} \coloneqq l^{-1}(p) \cap H_R$ for each \(p \in H_L\) (we have $\mathcal{S}_R^{p} \cap \mathcal{S}_R^{p'} = \emptyset$ for any distinct \(p,p' \in H_L\)) and then construct a matching $M'_{\rm star}$. The detailed construction of the matching $M'_{\rm star}$ is described in the following algorithm. 

\vspace{0.1cm}
\noindent \textbf{Algorithm C:}
\vspace{-0.1cm}
\begin{center}
\begin{adjustwidth}{0cm}{}
\begin{tabular}{p{1.4cm} p{10.6cm}}
    \textbf{Input:} & Pairwise disjoint sets of nodes $G_L,G_R,H_L,H_R$ such that $|G_L|=|H_L|$ and $|G_R|=|H_R|$, a set $\mathcal{S}_R^{p} \subseteq H_R$ for each $p \in H_L$ such that $\mathcal{S}_R^{p} \cap \mathcal{S}_R^{p'} = \emptyset$ for all $p \neq p'$, $x_{up} \in [0,1]$ for each $u \in G_L$ and $p \in H_L $ such that \\
    & \quad $\displaystyle\sum_{u \in H_L}x_{up} \leq 1$ for all $p \in H_L$ and $\displaystyle\sum_{p \in H_L}x_{up} \leq 1$ for all $u \in G_L$,  \\
    & $x_{vq} \in [0,1]$ for each $v \in G_R$ and $q \in H_R $ such that \\
    & \quad $\displaystyle\sum_{v \in G_R}x_{vq} \leq 1$ for all $q \in H_R$ and $\displaystyle\sum_{q \in H_R}x_{vq} \leq 1$ for all $v \in G_R$,  \\
    & $y_{upvq} \in [0,1]$ for each $u \in G_L,v \in G_R, p \in H_L$ and $q \in H_R$ such that\\
    & \quad $\displaystyle\sum_{u \in G_L } y_{upvq} \leq x_{vq}$ for all $v \in G_R,p \in H_L $ and $q \in H_R$, \\
    & \quad $\displaystyle\sum_{p \in H_L } y_{upvq} \leq x_{vq}$ for all $u \in G_L, v \in G_R$ and  $q \in H_R$, \\
    & \quad $\displaystyle\sum_{v \in G_R} y_{upvq} \leq x_{up}$ for all $u \in G_L,p \in H_L$ and $q \in H_R$, \\
    & \quad $\displaystyle\sum_{q \in H_R} y_{upvq} \leq x_{up}$ for all $u \in G_L, v \in G_R$ and  $p \in H_L$ \\
    \textbf{Output:} & Two matchings: $M_L$ between $G_L$ and $H_L$ and $M'_{\rm star}$ between $G_R$ and $H_R$
\end{tabular}
\end{adjustwidth}
\end{center}
\vspace{-0.1cm}
\begin{compactenum}
    \item For each $p \in H_L$, choose a node from $G_L \cup \{ \perp \}$ with probability $x_{up}$ for $u \in G_L$ and $1-\sum_{u \in G_L}x_{up}$ for $\perp$. Then, define a function $\pi : H_L \to G_L \cup \{\perp\}$ such that $\pi(p)$ is the chosen node for each $p$.
    \item For each $u \in G_L$, let $\pi^{-1}(u)$ be the set of node $p \in H_L$ such that $\pi(p) = u$. For each $u \in G_L$ such that $\pi^{-1}(u) \neq \emptyset$, randomly choose an element $p$ in $\pi^{-1}(u)$ and assign $(u,p)$ to a matching $M_L$.
    \item For each $p \in H_L$, $q \in \mathcal{S}_R^{p}$, and $v \in G_R$, let $ z^{p}_{vq} = y_{\pi(p) pvq}/x_{\pi (p) p}$ if $\pi (p) \neq \perp$, and $z^p_{vq}=0$ otherwise. 
    \item For each $p \in H_L$, consider $Z_p = \big[ z^p_{vq} \big]_{v \in G_R, q \in \mathcal{S}_R^{p}}$ as a fractional matching between $G_R$ and $\mathcal{S}_R^{p}$ (Lemma \ref{lem:prestar}). Let $\mathcal{M}_p$ be the set of all matchings between $G_R$ and $\mathcal{S}_R^{p}$. Then $Z_p$ can be written as a convex combination of matchings in $\mathcal{M}_p$ (Theorem 7.1.2 of \cite{lovasz2009matching}), i.e., $Z_p = \sum_{M \in \mathcal{M}_p} \alpha_M V_M$ where $V_M$ is a matrix representation, i.e., the incident (biadjacency) matrix, of $M \in \mathcal{M}_p$: $\big[ \delta^M_{vq} \big]_{v \in G_R, q \in \mathcal{S}_R^{p}}$ where $\delta^M_{vq} = 1$ if $(v,q) \in M$ and $\delta^M_{vq} = 0$ otherwise; and $\alpha_M \in [0,1]$ such that $ \sum_{M \in \mathcal{M}_p} \alpha_M \leq 1$. Choose one of those matchings with the probability equals to its $\alpha_M$. Then, define a function $\tau_p: \mathcal{S}_R^{p} \to G_R \cup \{\perp\}$ following the chosen matching $M$ such that $\tau_p(q) = v$ if $(v,q) \in M$ and $\tau_p(q) = \perp$ if $q$ is not incident to any edges in $M$. Note that $\operatorname{dom}(\tau_p) \cap \operatorname{dom}(\tau_{p'}) = \emptyset$ for any $p \neq p'$.
    \item Define a function $\tau: H_R \to G_R \cup \{\perp\}$ such that for each $q \in H_R$, $\tau(q)=\tau_{p}(q)$ if $q \in \mathcal{S}_R^{p}$ for some $p \in H_L$ and $\tau(q)=\perp$ otherwise.
    \item For each $v \in G_R$, let $\tau^{-1}(v)$ be the set of node $q \in H_R$ such that $\tau(q)=v$. For each $v \in G_R$ such that $\tau^{-1}(v) \neq \emptyset$, randomly choose a node $q \in \tau^{-1}(v)$ and assign $(v,q)$ to the matching $M'_{\rm star}$.
    \item Return $M_L$ and $M'_{\rm star}$.\\
\end{compactenum}

We first establish the following lemma: the matrix
\(Z_p=\big[z^p_{vq}\big]_{v\in G_R,\; q\in \mathcal{S}_R^{p}}\) arising in the
construction is a fractional matching (i.e., every row and column sum is at most \(1\)).
We then invoke this property to carry out Step~3 of the construction.

\begin{lemma} \label{lem:prestar}
    Given the input of Algorithm C, for each $p \in H_L$, the matrix $Z_p = \big[ z^p_{vq} \big]_{v \in G_R, q \in \mathcal{S}_R^{p}}$ constructed following Algorithm C is a fractional matching between $G_R$ and $\mathcal{S}_R^{p}$.
\end{lemma}

\begin{proof}
    Let $p \in H_L$. Following the condition of the input and the definition of $z^p_{vq}$,  $$\sum\limits_{q \in \mathcal{S}_R^{p}}z^{p}_{vq} = \sum\limits_{q \in \mathcal{S}_R^{p}} \frac{y_{\pi(p) pvq}}{x_{\pi (p) p}} \leq \sum\limits_{q \in  H_R} \frac{y_{\pi(p) pvq}}{x_{\pi (p) p}} \leq \frac{x_{\pi (p) p}}{x_{\pi (p) p}} = 1$$ and $$\sum\limits_{v \in G_R} z^{p}_{vq} = \sum\limits_{v \in G_R} \frac{y_{\pi(p) pvq}}{x_{\pi (p) p}} \leq \frac{x_{\pi (p) p}}{x_{\pi (p) p}} = 1$$ which imply that $Z_p = \big[ z^p_{vq} \big]_{v \in G_R, q \in \mathcal{S}_R^{p}}$ is a fractional matching between $G_R$ and $\mathcal{S}_R^{p}$. \qed
\end{proof}

The following lemma is leveraged to demonstrate Lemma \ref{lem:alg1.2}.

\begin{lemma} [Lemma 5.3 of \cite{nguyen2021maximum}] \label{lem:ind2}
    Let $S$ be a random subset of a set $V$ and $T$ be a random subset of a set $W$, and let $(v,w) \in V \times W$. Assume that for all $s,s' \in V$ and $t,t' \in W$ such that $s \neq s'$ and $t \neq t'$, the event that $s \in S$ and the event that $t \in T$ are independent to the event that $(s',t') \in S \times T$. If we pick an element from $S \times T$ at random, the probability of having $(v,w)$ as the chosen element is at least $\Pr[(v,w) \in S \times T]/\bigl( (\EX[|S|]+1) \cdot (\EX[|T|]+1) \bigr)$.
\end{lemma}

The following lemma follows from Lemma \ref{lem:ind2}.

\begin{lemma} \label{lem:star}
    If we proceed according to Algorithm C with the required input, then for $u \in G_L, p \in H_L, v \in G_R, q \in \mathcal{S}_R^{p}$, we have 
    $\Pr[(u,p) \in M_L \text{ and } (v,q) \in M'_{\rm star}] \geq y_{upvq}/4$.
\end{lemma}

\begin{proof}
    Let $p \in H_L$. By Lemma \ref{lem:prestar} and Theorem 7.1.2 of \cite{lovasz2009matching}, we know that $Z_p = \big[ z^p_{vq} \big]_{v \in G_R, q \in \mathcal{S}_R^{p}}$ is a convex combination of matchings between $G_R$ and $\mathcal{S}_R^{p}$. Let $\mathcal{M}_p$ denote the set of all matchings $M$ between $G_R$ and $\mathcal{S}_R^{p}$, each written in a matrix representation: $ V_M = \big[ \delta^M_{vq} \big]_{v \in G_R, q \in \mathcal{S}_R^{p}}$ where $\delta^M_{vq} = 1$ if $(v,q) \in M$ and $\delta^M_{vq} = 0$ otherwise. We then have $Z_p = \sum_{M \in \mathcal{M}_p} \alpha_M V_M$ where $\alpha_M$ is a nonnegative real number, and $ \sum_{M \in \mathcal{M}_p} \alpha_M \leq 1$. Hence, for each $v \in G_R$ and $q \in \mathcal{S}_R^{p}$, we have 
    \begin{align*}
        \Pr[\tau(q)=v|\pi(p)=u] &= \Pr[\tau_p(q)=v|\pi(p)=u]\\
        &= \sum_{M \in \mathcal{M}_p : (v,q) \in M}\alpha_M\\
        &= \sum_{M \in \mathcal{M}_p}\alpha_M \delta^M_{vq}\\
        &= z^p_{vq}.
    \end{align*}
    
    \noindent Thus, $\displaystyle \Pr[\tau(q)=v|\pi(p)=u] =y_{upvq}/x_{up}$ if $\pi(p) = u \in G_L$, and $\displaystyle \Pr[\tau(q)=v|\pi(p)=\perp] = 0$ by the definition of $z^p_{vq}$. Therefore, for each $v \in G_R$,
\begin{align*}
    \EX[|\tau^{-1}(v)|] &= \sum_{p \in H_L} \sum_{q \in \mathcal{S}_R^{p}} \Pr[\tau(q)=v]\\
    &= \sum_{p \in H_L} \sum_{q \in \mathcal{S}_R^{p}} \sum_{u \in G_L} \Pr[\tau(q)=v|\pi(p)=u] \cdot \Pr[\pi(p)=u]\\
    &= \sum_{p \in H_L} \sum_{q \in \mathcal{S}_R^{p}} \sum_{u \in G_L} \frac{y_{upvq}}{x_{up}} \cdot x_{up}\\
    &= \sum_{p \in H_L} \sum_{q \in \mathcal{S}_R^{p}} \sum_{u \in G_L} y_{upvq}\\
    &= \sum_{q \in \bigcup_{p \in H_L} \mathcal{S}_R^{p}} \sum_{u \in G_L} y_{up_qvq}\\
    &\leq \sum_{q \in H_R} x_{vq}\\
    &\leq 1
\end{align*}

\noindent where $p_q$ is a node $p \in H_L$ whose $\mathcal{S}_R^{p}$ contains $q$.  Recall that, in the proof of Lemma \ref{lem:rand}, for each $u \in G_L$, we have $\EX[|\pi^{-1}(u)|] \leq 1$ for this construction of $\pi$. Moreover, for $p,p' \in H_L$ and $q,q' \in H_R$ such that $p \neq p'$ and $q \neq q'$, the event that $p \in \pi^{-1}(u)$ and the event that $q \in \tau^{-1}(v)$ are independent to the event that $(p',q') \in \pi^{-1}(u) \times \tau^{-1}(v)$. Hence, given that $u \in G_L, p \in H_L, v \in G_R, q \in \mathcal{S}_R^{p}$, by Lemma \ref{lem:ind2}, we then get
\begin{align*}
    &\Pr[(u,p)\in M_L \text{ and } (v,q)\in M'_{\rm star}] \\
    &\qquad \geq \displaystyle \frac{\Pr[p \in \pi^{-1}(u) \text{ and } q \in \tau^{-1}(v)]}{(\EX[|\pi^{-1}(u)|]+1)(\EX[|\tau^{-1}(v)|]+1)}\\
    &\qquad \geq \displaystyle \frac{1}{4} \cdot \Pr[p \in \pi^{-1}(u)] \cdot \Pr[q \in \tau^{-1}(v)|p \in \pi^{-1}(u)]\\
    &\qquad = \displaystyle \frac{1}{4} \cdot \Pr[\pi(p) = u] \cdot \Pr[\tau(q) = v |\pi(p) = u]\\
    &\qquad = \displaystyle \frac{1}{4} \cdot x_{up} \cdot \frac{y_{upvq}}{x_{up}}\\
    &\qquad = \frac{y_{upvq}}{4}.
\end{align*} \qed
\end{proof}

We are now ready to formally prove Lemma \ref{lem:alg1.2}.

\begin{proof} [Proof of Lemma \ref{lem:alg1.2}]
Given $G_L, H_L, G_R, H_R, x^*_{up}$, and $y^*_{upvq}$. Construct for each \(p \in H_L\) a set $\mathcal{S}_R^{p}$ and proceed according to Algorithm C using the analogous input while $x_{up} = x^*_{up}$ and $y_{upvq} = y^*_{upvq}$. By Lemma \ref{lem:star}, we have
\[\Pr[(u,p) \in M_L \text{ and } (v,q) \in M'_{\rm star}|u \in G_L, p \in H_L, v \in G_R, q \in \mathcal{S}_R^p] \geq \frac{y^*_{upvq}}{4},\]

\noindent and thus, by using Lemma \ref{lem:withoutrandomG_L} on the fact that the event \say{$(u,p) \in M_L \text{ and } (v,q) \in M'_{\rm star}$ when $p=l(q)$} occurs only if $u \in G_L, p \in H_L, v \in G_R, q \in H_R$, we obtain \[\Pr[(u,p)\in M_L \text{ and } (v,q)\in M'_{\rm star}|p=l(q)] \geq \frac{y^*_{upvq}}{4} \cdot \frac{1}{32} = \frac{y^*_{upvq}}{128}.\]

\noindent From the way that the function $l$ is defined, we can conclude that $\Pr[(u,p)\in M_L \text{ and } (v,q)\in M'_{\rm star}] = y^*_{upvq}/128$ if $p=l(q)$, and $\Pr[(u,p)\in M_L \text{ and } (v,q)\in M'_{\rm star}] = 0$ otherwise. Recall that by this construction, $M_L$ is compatible. On the other hand, by the constraints of the linear programming, we have $y^*_{upvq}=0$ if $q \notin \mathcal{L}(v)$. Thus, for each $p \in H_L$, $v \in G_R$ and $q \notin \mathcal{L}(v) \cap \mathcal{S}_R^p$, we have
 $ \sum_{M \in \mathcal{M}_p} \alpha_M \delta^M_{vq} = z^p_{vq} = 0$.
 Hence, for any matching $M$ between $G_R$ and $\mathcal{S}_R^p$ that contains $(v,q) \notin E_{GH}$, $\alpha_M$ must be $0$. This implies that the matching $M'_{\rm star}$ is compatible. Therefore, we can consider the probability of each $(v,q) \in V_G \times V_H$ as the probability given above. With $|\mathcal{W}_q|\le \frac{\sqrt{n+k^2}+k}{2}+1$, we then have

\begin{align*}
    &\EX[Obj'(M_L,M_R)] \\
    &\qquad \geq \EX[Obj'(M_L,M'_{\rm star})]\\
    &\qquad= \sum_{u,v \in V_G}\sum_{p,q \in V_H} \Pr[(u,p) \in M_L \text{ and } (v,q) \in M'_{\rm star}] w_G(u,v)w_H(p,q)\\
    &\qquad= \sum_{q \in V_H:p=l(q)} \sum_{u,v \in V_G} \Pr[(u,p) \in M_L \text{ and } (v,q) \in M'_{\rm star}] w_G(u,v)w_H(p,q)\\
    &\qquad\geq \sum_{q \in V_H:p=l(q)} \sum_{u,v \in V_G} \frac{y^*_{upvq}}{128} w_G(u,v)w_H(p,q)\\
    &\qquad= \frac{1}{128} \sum_{q \in V_H} \sum_{u,v \in V_G} y^*_{ul(q)vq} w_G(u,v)w_H(p,q)\\
    &\qquad= \frac{1}{128} \sum_{q \in V_H} \max_{p \in V_H} \sum_{u,v \in V_G} y^*_{upvq} w_G(u,v)w_H(p,q)\\
    &\qquad\geq \frac{1}{128} \sum_{q \in V_H} \frac{1}{|\mathcal{W}_q|} \sum_{p \in \mathcal{W}_q} \sum_{u,v \in V_G} y^*_{upvq} w_G(u,v)w_H(p,q)\\
    &\qquad \geq \frac{1}{128} \sum_{q \in V_H} \frac{2}{\sqrt{n+k^2}+k+2} \sum_{p \in \mathcal{W}_q} \sum_{u,v \in V_G} y^*_{upvq} w_G(u,v)w_H(p,q)\\
    &\qquad= \frac{1}{64(\sqrt{n+k^2}+k+2)} \sum_{u,v \in V_G} \sum_{\substack{q \in V_H \\ p \in \mathcal{W}_q}} y^*_{upvq} w_G(u,v)w_H(p,q)\\
    &\qquad= \frac{1}{64(\sqrt{n+k^2}+k+2)} \cdot LP^*_1.
\end{align*}

\noindent Note that $\sqrt{n+k^2}+k+2 \leq \sqrt{n}+k +k+2 = O(\sqrt{n}+k)$. Therefore, we can conclude that $$\EX[Obj'(M_L,M_R)] \geq \frac{LP^*_1}{64(\sqrt{n+k^2}+k+2)}\geq \frac{LP^*_1}{O(\sqrt{n}+k)}.$$ \qed
\end{proof}

\section{Proof of Lemma \ref{lem:alg4.1}}
\label{appendix:lemma3}
Lemma \ref{lem:alg4.1} can be demonstrated as follows:

\begin{proof}[Proof of \ref{lem:alg4.1}]
Let $G_L, H_L, G_R, H_R$ and $x^*_{up}$ be those obtained in Steps~1-2 of Algorithm~1, and let $M_L'^{(i)}$ be the matching produced in Steps~3-5 of case $i$. Obtain $b$ random matchings between $G_R$ and $H_R$, denoted by $M_{\rm rand}'^{(1)},\ldots,M_{\rm rand}'^{(b)}$. Define
\[
\leftindex^{b}{M}'_{L} \coloneqq \bigcup_{i=1}^{b} M_{L}'^{(i)}
\qquad\text{and}\qquad
\leftindex^{b}{M}'_{\rm rand} \coloneqq \bigcup_{i=1}^{b} M_{\rm rand}'^{(i)}.
\]

 Lemma~\ref{lem:rand} provides us \[\Pr[(u,p) \in M'^{(i)}_L|u \in G_L, p \in H_L] \geq \displaystyle\frac{x^*_{up}}{2b}\]
\noindent and \[\Pr[(v,q) \in M'^{(i)}_{\rm rand}|v \in G_R, q \in H_R] = \displaystyle\frac{1}{|G_R|} \geq \frac{2}{n}\]
\noindent for all $i=1,2,...,b$.  Note that the collection of $2b$ events \say{$(u,p) \in M'^{(i)}_L$} and \say{$(v,q) \in M'^{(i)}_{\rm rand}$} above are pairwise independent. Hence, given that $u \in G_L, p \in H_L, v \in G_R,$ and $ q \in H_R$, we have
\begin{align*}
    & \Pr[(u,p) \in \leftindex^b{M}'_L \text{ and } (v,q) \in \leftindex^b{M}'_{\rm rand}] \\
    & \quad\quad = \Pr[(u,p) \in \leftindex^b{M}'_L] \Pr[(v,q) \in \leftindex^b{M}'_{\rm rand}]\\
    & \quad\quad = \bigl( 1 - \Pr[(u,p) \notin \leftindex^b{M}'_L] \bigr)  \bigl( 1 - \Pr[(v,q) \notin \leftindex^b{M}'_{\rm rand}] \bigr)\\
    & \quad\quad = \bigl( 1 - \prod_{i=1}^b\Pr[(u,p) \notin M'^{(i)}_L] \bigr)  \bigl( 1 - \prod_{i=1}^b\Pr[(v,q) \notin M'^{(i)}_{\rm rand}] \bigr)\\
    & \quad\quad \geq \bigl( 1 - (1-\frac{x^*_{up}}{2b})^b \bigr)  \bigl( 1 - (1 - \frac{2}{n})^b \bigr)\\
    & \quad\quad \geq \bigl( 1 - (e^{-\frac{x^*_{up}}{2b}})^b \bigr)  \bigl( 1 - (e^{- \frac{2}{n}})^b \bigr)\\
    & \quad\quad = (1-e^{-\frac{x^*_{up}}{2}})(1-e^{- \frac{2b}{n}})\\
    & \quad\quad \geq \Bigl( 1 - \frac{1}{e} \Bigr)^2 \cdot \frac{x^*_{up}}{2} \cdot \frac{2b}{n}\\
    & \quad\quad = \Bigl( 1 - \frac{1}{e} \Bigr)^2 \cdot \frac{bx^*_{up}}{n}.
\end{align*}
\noindent The last two inequalities follow from the fact that $1+x \leq e^x$ for all $x \in \mathbb{R}$ and $1-e^{-x} \geq (1-1/e)x$ for $0 \leq x \leq 1$. Hence, by Lemma \ref{lem:withoutrandomG_L}, we then have that for any $u,v \in V_G$ and $p,q \in V_H$, 
\[\Pr[(u,v) \in \leftindex^b{M}'_L \text{ and } (v,q) \in \leftindex^b{M}'_{\rm rand}] \geq  \Bigl( 1 - \frac{1}{e} \Bigr)^2 \cdot \frac{bx^*_{up}}{n} \cdot \frac{1}{32} = \frac{bx^*_{up}}{c_1n}.\]

\noindent where $c_1 = 32/(1-1/e)^2$. By the way of choosing $\leftindex^b{M}_R$ in Algorithm 2, for a corresponding $\leftindex^b{M}'_L$ and $\leftindex^b{M}_L$, we know that $Obj'(\leftindex^b{M}_L,\leftindex^b{M}_R) \geq Obj'(\leftindex^b{M}'_L,\leftindex^b{M}'_{\rm rand})$. Thus, $\EX[Obj'(\leftindex^b{M}_L,\leftindex^b{M}_R)] \geq \EX[Obj'(\leftindex^b{M}'_L,\leftindex^b{M}'_{\rm rand})]$. Furthermore, by a similar argument to which used for Lemma~\ref{lem:alg1.1} together with the fact that $|\mathcal{U}_p| \geq \sqrt{n}/\sqrt{b}$ and $\sum_{q \in V_H} y_{upvq} \leq bx_{up}$ for all $u,v \in V_G$ and $p \in V_H$, we have
\begin{align*}
    \EX[Obj'(\leftindex^b{M}'_L,\leftindex^b{M}'_{\rm rand})]  &\geq \frac{b}{c_1n} \cdot \frac{\sqrt{n}}{\sqrt{b}} \cdot \frac{1}{b} \sum_{u,v \in V_G} \sum_{\substack{p \in V_H \\ q \in V_H \backslash \mathcal{U}_p}} w_G(u,v)w_H(p,q)y^*_{upvq}\\
    & = \frac{1}{c_1\sqrt{bn}} \sum_{u,v \in V_G} \sum_{\substack{p \in V_H \\ q \in V_H \backslash \mathcal{U}_p}} w_G(u,v)w_H(p,q)y^*_{upvq}
\end{align*}
\noindent This implies $\EX[Obj'(\leftindex^b{M}_L,\leftindex^b{M}_R)] \geq (LP^*_2/(c_1\sqrt{bn})) = LP^*_2/O(\sqrt{bn})$. \qed
\end{proof}

\section{Proof of Lemma \ref{lem:alg4.2}}

\label{appendix:lemma4}

Lemma \ref{lem:alg4.2} can be demonstrated as follows:

\begin{proof}[Proof of Lemma \ref{lem:alg4.2}]
Recall the function $l$ defined in the previous section. Given $G_L, H_L, G_R, \allowbreak H_R$, and the solutions $x^*_{up}$ and $y^*_{upvq}$ from Steps~1-2 of Algorithm~2, define for each $p\in V_H$
\[
\mathcal{S}_R^{p} \coloneqq l^{-1}(p)\cap H_R .
\]
Proceed to execute Algorithm~C for $b$ iterations on the corresponding sets, using
\[
x_{up} \coloneqq \frac{x^*_{up}}{b}
\qquad\text{and}\qquad
y_{upvq} \coloneqq \frac{y^*_{upvq}}{b^{2}}
\]
for the relevant nodes. By the constraints of Relaxed~LP2, these scaled values $x_{up}$ and $y_{upvq}$ satisfy the feasibility requirements of Algorithm~C. Let $M_L'^{(i)}$ and $M_{\rm star}'^{(i)}$ be the matchings produced in the $i$-th iteration, and define
\[
\leftindex^{b}{M}'_{L} \coloneqq \bigcup_{i=1}^{b} M_{L}'^{(i)}
\qquad\text{and}\qquad
\leftindex^{b}{M}'_{\rm star} \coloneqq \bigcup_{i=1}^{b} M_{\rm star}'^{(i)} .
\]

    Following Lemma~\ref{lem:star}, we have 
    \[\Pr[(u,p) \in M'^{(i)}_L \text{ and } (v,q) \in M'^{(i)}_{\rm star}|u \in G_L, p \in H_L, v \in G_R, q \in \mathcal{S}_R^p] \geq \frac{y^*_{upvq}}{4b^2}\]
    for all $i=1,2,...,b$. Since each iteration is independent from each other, the collection of the events \say{$(u,p) \in M'^{(i)}_L \text{ and } (v,q) \in M'^{(i)}_{\rm star}$} is pairwise independent. Thus, given that $u \in G_L, p \in H_L, v \in G_R,$ and $ q \in \mathcal{S}_R^p$, we have
\begin{align*}
    & \Pr[(u,p) \in \leftindex^b{M}'_L \text{ and } (v,q) \in \leftindex^b{M}'_{\rm star}] \\
    & \quad\quad \geq \Pr[\bigl( (u,p) \in M'^{(1)}_L \text{ and } (v,q) \in M'^{(1)}_{\rm star} \bigr) \text{ or } \\ &\qquad\qquad\quad\ \dots \text{ or } \bigl( (u,p) \in M'^{(b)}_L \text{ and } (v,q) \in M'^{(b)}_{\rm star} \bigr)] \\
    & \quad\quad = 1 - \Pr[\bigl( (u,p) \notin M'^{(1)}_L \text{ or } (v,q) \notin M'^{(1)}_{\rm star} \bigr) \text{ and } \\ &\qquad\qquad\qquad\quad \dots \text{ and } \bigl( (u,p) \notin M'^{(b)}_L \text{ or } (v,q) \notin M'^{(b)}_{\rm star} \bigr)] \\
    & \quad\quad = 1 - \prod_{i=1}^b\Pr[(u,p) \notin M'^{(i)}_L \text{ or } (v,q) \notin M'^{(i)}_{\rm star}]\\
    & \quad\quad \geq 1 - (1-\frac{y^*_{upvq}}{4b^2})^b \\
    & \quad\quad \geq 1 - (e^{-\frac{y^*_{upvq}}{4b^2}})^b \\
    & \quad\quad = 1-e^{-\frac{y^*_{upvq}}{4b}}\\
    & \quad\quad \geq \Bigl( 1 - \frac{1}{e} \Bigr) \cdot \frac{y^*_{upvq}}{4b}.
\end{align*}
Then by Lemma \ref{lem:withoutrandomG_L}, for any $u,v \in V_G$ and $p,q \in V_H$,
\[
\Pr\!\bigl[(u,p)\in \leftindex^{b}{M}'_{L}\ \text{and}\ (v,q)\in \leftindex^{b}{M}'_{\rm star}\,\bigm|\, p=l(q)\bigr]
\;\ge\;
\Bigl(1-\tfrac{1}{e}\Bigr)\cdot \frac{y^*_{upvq}}{4b}\cdot \frac{1}{32}
\;=\; \frac{y^*_{upvq}}{c_2\, b},
\]
where $c_2 \coloneqq 128/(1-1/e)$. Moreover,
\[
\Pr\!\bigl[(u,p)\in \leftindex^{b}{M}'_{L}\ \text{and}\ (v,q)\in \leftindex^{b}{M}'_{\rm star}\,\bigm|\, p\neq l(q)\bigr]=0,
\]
by the construction of $l$. Using the same reasoning as in the proof of Lemma~\ref{lem:alg4.1}, we obtain
\[
\EX[Obj'\!\bigl(\leftindex^{b}{M}_{L},\leftindex^{b}{M}_{R}\bigr)]
\;\ge\;
\EX[Obj'\!\bigl(\leftindex^{b}{M}'_{L},\leftindex^{b}{M}'_{\rm star}\bigr)]\!.
\]

Following the same way of proof provided for Lemma~\ref{lem:alg1.2} along with the fact that $1/|\mathcal{U}_q| \geq \sqrt{b}/(2\sqrt{n})$, we also have
\begin{align*}
    \EX[Obj'(\leftindex^b{M}'_L,\leftindex^b{M}'_{\rm star})] &\geq  \frac{1}{c_2b} \cdot \frac{\sqrt{b}}{2\sqrt{n}} \sum_{u,v \in V_G} \sum_{\substack{q \in V_H \\ p \in \mathcal{U}_q}} y^*_{upvq} w_G(u,v)w_H(p,q)\\
    & = \frac{1}{O(\sqrt{bn})} \sum_{u,v \in V_G} \sum_{\substack{q \in V_H \\ p \in \mathcal{U}_q}} y^*_{upvq} w_G(u,v)w_H(p,q)
\end{align*}
\noindent Thus, we obtain $\EX[Obj'(\leftindex^b{M}_L,\leftindex^b{M}_R)] \geq LP^*_1/O(\sqrt{bn})$ as desired. \qed
\end{proof}

\end{subappendices}

\end{document}